\documentclass[10pt]{article}
\usepackage{graphicx}
\usepackage{caption}
\usepackage{subcaption}

\begin{document}

\title{Particle Dark Matter in the 1980s and 1990s\footnote{To appear in the 4th International Symposium on the History of Particle Physics, to be published by Cambridge University Press.}}
  
\author
{Michael S. Turner \\
\normalsize Kavli Institute for Cosmological Physics\\
\normalsize University of Chicago, Chicago, IL  60637-1433\\
\\
\normalsize Department of Physics and Astronomy\\
\normalsize The University of California, Los Angeles\\
\normalsize Los Angeles, CA  90095-1547\\
\small 
email: mturner@uchicago.edu
\\
}
\date{}      
\maketitle

\vskip 0.2in

\begin{abstract}
{In 1980, the Universe was made of stars, cosmology was the province of less than 100 astronomers, the Hubble constant was only known to within a factor of two, and the hot big bang was the ``the standard model of cosmology.''  Particle physics was a thriving enterprise concerned with the inner space of quarks, leptons and the strong, weak and electromagnetic interactions.  The Standard Model was newly established and the aspirations were for a ``grand unification'' of the forces and particles.  By 2000, the agendas of inner space and outer space were inextricably linked, and $\Lambda$CDM with its inflation, particle dark matter and dark energy -- and a Hubble constant reliably known to better than 10\% -- was the new standard model.  The big questions connecting inner space and outer space included the origin of ordinary matter, the identity of dark matter, the nature of dark energy, and an understanding of the origin of space, time and the Universe.  Particle dark matter, the idea that the bulk of the mysterious dark matter is a new, long-lived or stable elementary particle, played a central in bringing together inner space and outer space, and today is still a hot topic in both fields.}

\end{abstract}
\vskip 0.1in

\noindent  

\section{Dark matter in astronomy}
Today we know that stars account for little of the mass/energy content of the Universe (about 0.5\%)  That was not always the case, and the path to this realization took a half century.

There were early indications that the Universe was more than meets the eye, that is, not all stars.    They include studies of the Andromeda galaxy by Lundmark in 1930 \cite{Lundmark} and Babcock in 1939 \cite{Babcock}; and Smith's analysis of the Virgo cluster in 1936 \cite{Smith}, all of which found, by measuring the motions of stars and galaxies in bound systems, the need for more gravitational mass than stars contributed.

The most well known work from this early period is that of Zwicky in 1933, who studied the Coma cluster of galaxies \cite{Zwicky}.  While the bulk of his paper was about his tired light hypothesis to explain cosmological redshifts, he determined the mass of Coma using the virial theorem and estimated that stars failed by more than a factor of 10 to account for the needed gravitational mass.  And, he coined the term dark matter.

In the 1930s cosmology was not yet even a field; less than a hundred galaxies had measured redshifts; and ``high redshift'' had not yet reached $z=0.1$.\footnote{Today, in the era of JWST, there are 100s of galaxies with measured redshifts greater than 10, with the current record holder at $z\simeq 15$.}  Even stars were not well understood.  While Eddington had proposed fusion as their energy source in 1920, Bethe's landmark paper, {\it Energy production in stars,} did not appear until 1939. Astronomy was not yet ready for dark matter.

Things changed around 1980.  In the late 1970s, using 21 cm emission from clouds of gas, radio astronomers measured the rotation curve of Andromeda beyond stars and found that it was not ``planetary,'' that is decreasing as $1/\sqrt{r}$ as expected if the stars were providing the necessary gravity to hold M31 together. The rotation velocity was constant at large distances from the center \cite{Roberts,Bosma}.  

Radio astronomers, who viewed themselves as physicists, and astronomers did not always communicate well; further, the radio technique had a wide aperture and the results were subject to interpretation.  In any case, the radio astronomers did not make the connection between flat rotation curves and the presence of dark matter.

From 1978 to 1988 Vera Rubin and her collaborator Kent Ford used his new instrumentation (electrostatic photomultiplier tubes) to make optical measurements of 100s of galactic rotation curves.  They established that ``flat'' rotation curves are the rule and not the exception, and that the masses of galaxies continued to rise with distance from a galaxy's center.  In an influential {\it Science} paper published in 1983 Rubin made the case that spiral galaxies have dark matter halos, as well as linking it to the dark matter seen in clusters by Zwicky and others \cite{Rubin1983}.  In an important review, Faber and Gallagher \cite{FG1979} made the case that most of the mass detected in gravitationally-bound systems is dark. 

In 1985, the IAU made it ``official'' that dark matter is a thing in astronomy by holding a symposium on the subject at Princeton, {\it IAU Symposium \#117:  Dark Matter in the Universe}.  I contributed a 45-page paper describing the particle dark matter candidates \cite{Turner1985}.  It is also where I first met Vera Rubin.\footnote{Our paths crossed many times thereafter.  Vera did not think much of particle dark matter and was more sympathetic to MOND \cite{MOND}.  She was a strong advocate for women in astronomy, and when I was chair of the astronomy department at UChicago she sent me a postcard saying that if she were a bright young woman she would not come to UChicago for graduate school because there were no women on the faculty -- that changed under my watch.  As President of the APS in 2013, I participated in the celebration honoring her and Kent Ford when the APS made Carnegie's Department of Terrestrial Magnetism a physics historical site because of their dark matter work.  In preparing to give the summary talk at her memorial conference in 2019 at Georgetown University, I came to better understand and appreciate her critical role in dark matter, at time when a number of individuals were down playing it; see, e.g., Ref.~\cite{Tremaine2017}.}

\section{Particle dark matter}
With the exception of Steven Weinberg and his influential books, {\it Gravitation and Cosmology} \cite{Weinberg1972} and {\it The First Three Minutes} \cite{Weinberg1978}, particle physicists in the 1970s were not too interested in a field where it was said the errors are in the exponents and flux is measured in magnitudes.  Nonetheless, his text laid out the hot big bang model and the evidence for it, as naming it, ``the standard model of cosmology,'' before that moniker was used for the Standard Model of particle physics.

In the 1970s the hot big bang model was sufficiently well-established that the Universe could be used as a ``heavenly laboratory.''  Cowsik and McClelland \cite{CM1972} and Marx and Szalay \cite{MS1975} used an upper limit to the present mass density and the relic abundance of neutrinos to constrain neutrino masses to the tens of eV range.  In 1977, Lee and Weinberg \cite{LW1977} computed the relic abundance of a hypothetical, long-lived massive ($m \gg \,$MeV) neutrino species, and used it to place a lower limit to its mass.  Their calculation would become the template for calculating relic abundances of dark matter candidates that were once in thermal equilibrium.

Another influential paper linking particle physics and cosmology was that of Steigman, Schramm and Gunn \cite{SSG1977}.  They used the big bang production of $^4$He to constrain the number of light neutrino species to be less than or equal to 5.  With better data on the primordial $^4$He abundance and constrains to the baryon density, the limit would improve to less than 3.4 \cite{OSSW1990} at the time the first SLC and LEP measurements of the width of the $Z^0$ boson constrained the number of neutrino species to be less than 4.  This was a milestone:  a heavenly laboratory ``prediction'' was confirmed in an earthly laboratory.

\subsection{GUTs and the neutrino-dominated Universe}
In 1980 a shift occurred with the emergence of the neutrino-dominated Universe.  Cosmology was no longer just in service to particle physics, but particle physics had something to offer cosmology:  a dark matter candidate.

Four results in 1980 paved the way:  two theoretical and two experimental.  The two theoretical papers built upon the excitement around the idea of Grand Unified Theories (GUTs) and their prediction of proton decay, neutrino mass and oscillation, and baryogenesis.\footnote{In my opinion, the idea of grand unification got short shrift at this Symposium, perhaps because only its prediction of neutrino mass and oscillation have been experimentally verified.  It inspired theorists to think about the connections between particle physics and cosmology, beginning with baryogenesis and inflation, and it motivated experimentalists to initiate a host of non-accelerator particle physics experiments, including proton decay, neutrino oscillation, magnetic monopole searches, and dark matter detection.}  As it turned out, both experimental papers were wrong, and we now know that neutrinos contribute only a tiny amount to the cosmic mass/energy budget, about 0.15\%, and comprise a small fraction of the dark matter density.  Nonetheless, they are a proof of principle for particle dark matter.

The first theoretical result was the see-saw mechanism \cite{SeeSaw}, which provided a theory for how neutrinos could have masses much smaller than those of their charged leptonic partners and predicted eV level masses for neutrinos, which could be tested in neutrino oscillation experiments.  The second theoretical result was the April 1980 paper by Schramm and Steigman entitled, {\it A neutrino-dominated Universe} \cite{SS1980}.  It won first place that year in the Gravity Research Foundation's annual essay competition.\footnote{I believe it was much more influential than the 22 citations it has garnered because of the proselytizing done by its influential authors.  For astronomers and particle physicists, {\it General Relativity and Gravitation} is not an oft-read journal; the {\it Ap.J.} version of the paper did garner 122 citations.}

The two wrong experiments:  Fred Reines's paper announcing the discovery of neutrino oscillations from his experiment at the Savannah River reactor \cite{Rei1980} with a mass difference squared of order a few eV and Lubimov's paper announcing the discovery of neutrino mass in a tritium beta decay experiment \cite{Lubimovetal1980}, $14\,{\rm eV} < m_\nu < 46\,$eV (99\% confidence).  

Particle dark matter was off to the races with the neutrino-dominated Universe.  In this picture, structure forms from the top down (Zel'dovich's pancake theory); that is, large structures -- superclusters -- form first and then fragment into smaller structures -- galaxy clusters and galaxies.  

From my perspective, as cosmology noobie at the time, having a new idea to pursue and test revitalized the study of structure formation.  While the gravitational instability theory of structure formation was firmly in place and the physics of a baryons-only Universe was well understood, there too many free parameters -- the total matter density, the spectrum of the density inhomogeneities, and their nature adiabatic or isothermal -- and not enough data.  Neutrinos solved the parameter problem, in a new and novel way.

\subsection{Nuffield rocket fuel}
In the gravitational instability picture of structure formation cosmic structures grow from small inhomogeneities in the matter distribution, at a rate regulated by the expansion of the Universe, which in turn is determined by the composition of the Universe.  By the end of the Nuffield Workshop at Cambridge University in July 1982, essentially all of the input parameters for structure formation had been determined.  I was a participant at Nuffield, and this workshop remains one of the highlights of my career.

Nuffield was a before/after moment for the current inflationary paradigm, that is,  slow-roll inflation with almost scale-invariant, adiabatic Gaussian perturbations created from quantum fluctuations \cite{QM_Perts}.   Guth introduced the idea of inflation in 1980 \cite{Guth1980}, but in the same paper showed conclusively that his first-order phase-transition model didn't work.  Around the time plans were being made for the Nuffield workshop, Linde \cite{Linde1982} and Albrecht and Steinhardt \cite{AS1982} introduced the ``slow-roll'' fix for inflation.  After reading Linde's preprint in December 1981, Wilczek and I began working out how reheating in the inflationary Universe took place, resulting in our paper with Albrecht and Steinhardt \cite{ASTW1982}.  

Leading up to the Nuffield meeting, Steinhardt and I, Starobinskii, Hawking, and Guth and Pi were all working on how slow-roll inflation could produced density perturbations from quantum fluctuations.  By the time of the meeting, Guth and Pi were still working, and the rest of us arrived at Nuffield with papers which featured different wrong results.  Remarkably, by the end of the meeting, we had all converged on the correct answer, under the watchful eyes of the participants at the meeting.\footnote{Bardeen, who was also at the meeting, was particularly interested in Steinhardt and my results because we were using his perturbation formalism.  He found an error in our integration of a key equation and he became our collaborator.  He reported the error to me at midnight before the morning of my talk!}

Inflation also predicts a flat, critical density Universe, thereby specifying the total mass/energy content of the Universe to be equal to the critical density.  The big-bang production of deuterium limited the baryonic contribution to be less than around 10\% (see \S\ref{baryons}).  That left the other 90\% of the Universe to be non-baryonic, at least if inflation was correct.  Voila:  the neutrino-dominated Universe.

For neutrinos, the spectrum of the density perturbations was not essential to know.  This is because the ``freestreaming'' of neutrinos, which were still relativistic when the Universe became matter-dominated and structure formation began, washes out perturbations on small scales \cite{BES1980} leading to the ``top down,'' Zel'dovich pancake scenario of structure formation, known as hot dark matter. 

For the other limiting possibility, dark matter particles that are very non-relativistic at the time structure formation began -- so called cold dark matter -- the spectrum is essential.  And inflation provided the necessary information -- almost scale-invariant, Gaussian, and adiabatic.\footnote{At the time, the term adiabatic was used for what are now known as curvature perturbations.  In a two-component Universe -- baryons and radiation -- curvature perturbations are adiabatic perturbations in both the baryons and photons; and isothermal -- noncurvature perturbations -- are only in the baryons during the radiation epoch.  Particle dark matter created the need for a new nomenclature.}

Another important development at the Nuffield workshop was birth of the first cold dark matter candidate,  the axion.  Wilczek \cite{Wilczek_Nuffield} describes in his contribution to the proceedings the realization at the meeting that the ``invisible axion'' with a GUT PQ-symmetry breaking scale, would contribution too much mass density to the Universe today,\footnote{Wilczek was working with John Preskill, another attendee at Nuffield and Mark Wise \cite{PWW}; other independent work around the same time came to the same conclusion \cite{AS1983,DF1982}.} and lowering the symmetry breaking scale to $10^{12}\,$GeV would lead to an axion-dominated Universe.  Wilczek, Zee and I wrote the first paper on cold dark matter in 1983 \cite{TWZ1983} -- although we did not get all the details right.

\subsection{SUSY dark matter}
During the 1980s and 1990s, low-energy supersymmetry (SUSY) was a dominant theme in theoretical particle physics (as described elsewhere in these proceedings).  The necessity of R-parity conservation to stabilize the proton, led to one of the most attractive dark matter candidates:  the lightest supersymmetric particle (LSP), which is often the lightest neutralino (a combination of the photino, zino and higgsino).  The LSP is stable (owing to R-parity conservation), has weak interactions with  SM particles, and mass of order 100 GeV to many TeV.  And, it turned out that the neutralino was an ideal dark matter candidate.

How the neutralino came to be the leading dark matter candidate is a complicated story.  In 1981, Pagels and Primack \cite{PP1981} wrote a paper about a low-energy supersymmetry model where a keV-mass gravitino was the LSP.  They  argued that relic gravitinos could be the primary constituent of the Universe, and that unlike neutrinos, structures as small as galaxies and small groups of galaxies could form from the initial density perturbations because their freestreaming length would be significantly smaller.  

As it turns out, the Pagels-Primack gravitino was the first example of the intermediate case now known as warm dark matter \cite{BST1982}:  when structure formation begins, keV gravitinos would have been mildly-relativistic.  Their SUSY model was superceded by the more familiar supersymmetric extensions of the SM \cite{Haber1993}, where the neutralino is the LSP.  

One of the first papers -- if not the first paper -- to address the cosmological consequences of low-energy SUSY models was that of Ellis, Hagelin, Nanopoulos, Olive and Srednicki (1984), who used a Lee-Weinberg analysis of the relic abundance of neutralinos to constrain their SUSY model \cite{EHNOS}.  They did not mention relic neutralinos as comprising the dark matter, but they were concerned about the impact of relic gravitinos on baryogenesis.  

Cosmology was still just being used in service to particle physics, with no quid pro quo.  Nonetheless, this paper marks the beginning of neutralino dark matter.

\subsection{Hot, warm and cold dark matter}
 
 It was now ``game on'' for structure formation:  inflationary scale-invariant density perturbations, a flat Universe, a few baryons, and three different possibilities for  dark matter, hot, warm and cold, with the first and the third getting the bulk of the attention.  The dark matter candidates were neutrinos (hot), axions and neutralinos (cold) and gravitinos (warm).

The time was also ripe to test these new ideas by comparing the present Universe with simulations of the cosmic structure that would form in a particle dark matter universe.  The state-of-the-art in simulating the Universe was 32,000 particles on a $64^3$ grid, with each particle representing a very massive galaxy with gravitational interactions only.  The state-of-the-art in characterizing the large-scale structure of the Universe then was the first Center for Astrophysics (CfA$_1$) redshift survey:  2400 galaxies with mean redshift $z = 0.016$ \cite{DHLT1982}.  

The combination was good enough to strongly disfavor hot dark matter and establish consistency with cold dark matter, but not good enough to distinguish between warm and cold dark matter or to determine cosmological parameters, e.g., $\Omega_0$ or $\Omega_M$.

For comparison, today the largest simulations have trillions of particles, some  incorporate the ``gastrophysics'' of the baryons to simulate how galaxies light up, and the largest redshift survey (DESI) has 39 million (and counting) redshifts out to $z=4$.  Moore's law in cosmology has enabled precision cosmology.

In 1984, cold dark matter became ascendant.  Peebles \cite{PJEP1984} laid out the basics of cold dark matter (CDM), and the first review of CDM appeared in {\it Nature} \cite{Blumenthal1984}.  Over the next 15 or so years, there would be issues for CDM, but it was the theory with at least nine lives -- it also bounced back.  An especially nagging issue was the value of $\Omega_M$: $\Omega_M = 0.2$ seemed to be a better fit to the data -- but as mentioned above, the data was not yet good enough for precision testing.

While the neutrino-dominated Universe only lasted a year or two, it ushered in a powerful new idea:  particle dark matter and a Universe whose infrastructure was created and held together by the gravity of elementary particles not atoms.

Cosmologists were quicker to take a particle dark-matter dominated Universe seriously.  Even after the Ellis et al paper \cite{EHNOS} that marks the birth of neutralino dark matter, most of the papers on SUSY dark matter used the relic abundance as a constraint to SUSY models rather than touting the neutralino as a dark matter candidate.  Of course, particle physicists were fully occupied with testing and understanding low-energy supersymmetry and it dominated the experimental agendas of LEP, the TeVatron and the LHC.

Beginning in the 1990s,  ``evidence'' for SUSY would include the neutralino as a cold dark matter candidate; see e.g., Ref.~\cite{Kane2000}, or Google AI which returns ``viable cold dark matter candidate'' as \#3 in the reasons for believing in low-energy supersymmetry.  And while I cannot find the origins of the term ``WIMP miracle,'' it is meant to enthusiastically describe the coincidence (hint?) that low-energy supersymmetry and the weak interactions of its SUSY particles leads to the neutralino having the right abundance to be the dark matter particle.

Today, particle theorists and experimentalists are all in on dark matter.  With the absence of evidence for the most compelling candidates, theorists are busy proposing new candidates, motivated primarily by cosmology and worrying about how they fit into particle physics later.  Experimentalists continue to pursue particle dark matter -- new and old candidates -- at accelerators, tabletop experiments, underground experiments and through the search for their annihilation products.

\subsection{Detecting dark matter}

By the mid 1980s, and certainly by 1990, the most compelling candidates for particle dark matter were neutrinos, neutralinos (or WIMPs\footnote{The term WIMP, for Weakly Interacting Massive Particle, was introduced by Steigman and me around this time \cite{ST1986}; originally, it was meant to refer to any particle dark matter candidate, but it came to refer to dark matter particles whose interactions with SM particles were ``weak interaction'' strength, the most notable example being the neutralino.}) and axions.  Neutrinos actually exist, you get three tries and there is theoretical motivation for the needed eV masses.  One very big strike against them:  they don't produce the observed structure seen in the Universe today!

Neutralinos are theoretically well-motivated for particle physics reasons, their relic abundance arises due to their incomplete annihilation in the early Universe, and their annihilation cross sections being weak-interaction strength gives them the right relic abundance to be the dark matter, the so-called WIMP miracle.\footnote{For a discussion of how this works, without the word miracle, see Ch.~5 of Ref.~\cite{KT1990}.}  Further, such an interaction cross section with SM particles makes them detectable in three different ways; see below.

The axion is perhaps even more theoretically compelling:  it is the best (known) solution to the strong CP problem, a major shortcoming of QCD.  They too are detectable.  Lacking is the equivalent of the WIMP miracle:  namely, a compelling theoretical reason for the PQ symmetry breaking scale being that which leads to axions accounting for the dark matter.

The ultimate compliment that an experimentalist can pay to a theorist is to take their idea seriously enough to test it.  There were a handful of key papers that pointed out how particle dark matter could be tested: 
\begin{itemize}
    \item Sikivie's 1983 paper which showed that the so-called invisible axion was actually visible through its electromagnetic couplings \cite{Sikivie1983}, including axions that could account for the dark matter.
    \item Goodman and Witten's 1985 paper that outlined how halo dark matter particles of mass greater than a few GeV could be detected by the energy they deposit in a detector through elastic scattering \cite{GoodmanWitten1985}.
    \item Silk and Srednicki's 1984 paper that pointed out that annihilation products (e.g., positrons, antiprotons, and gamma rays) of halo dark matter particles might be detectable \cite{SilkSrednicki1984}.
    \item Silk, Olive and Srednicki showed that high-energy neutrinos from dark-matter annihilations in the sun could be detectable \cite{SOS1985}.
    \item Freese \cite{Freese1986} and Krauss, Srednicki and Wilczek \cite{KSW1986} showed that neutrinos from dark-matter annihilations in the Earth might be detectable as well.
\end{itemize}

\begin{figure}[tbp]
\center\includegraphics[width = 0.8\textwidth]{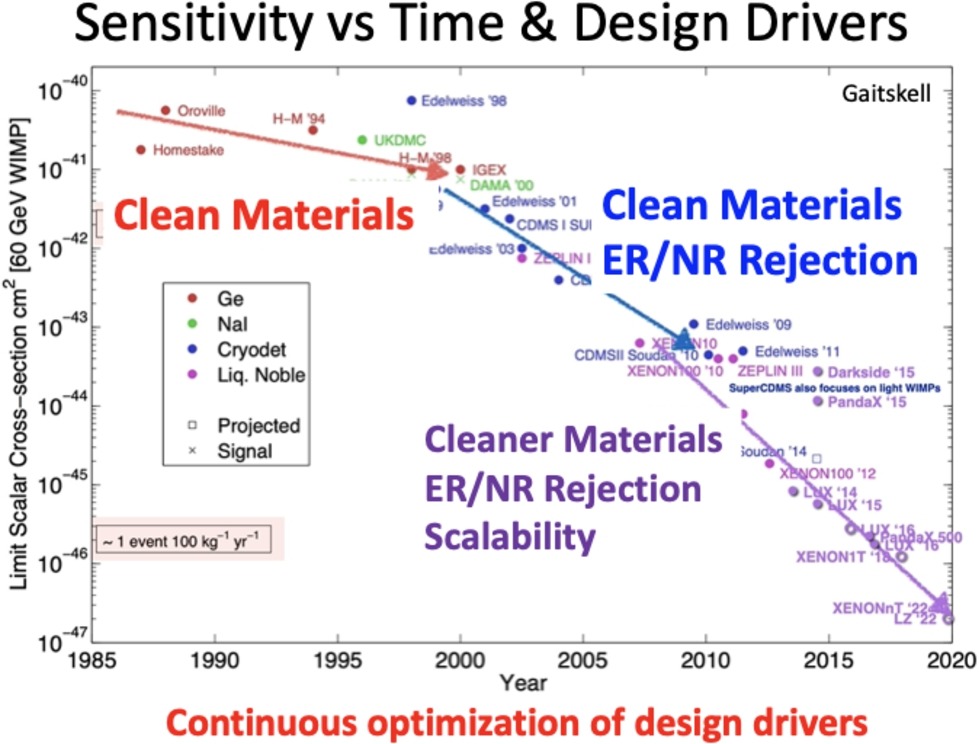}
\caption{Increasing sensitivity of direct detection experiments from the two first experiments in 1988 and until 2020.}
\label{Direct_progress}     
\end{figure}

In 1988 the first review of dark-matter detection appeared \cite{PSS1988}, and the first results (negative) for the search for dark matter, from double-beta decay experiments re-purposed for dark-matter detection, were reported \cite{Ahlen1987,Caldwell1988}.  
Since that time, the quest to detect and identify the dark-matter particle directly (halo dark-matter particles), indirectly through their annihilation products, and through production at accelerators has occupied particle physicists.  Fig.~\ref{Direct_progress} shows the Moore's Law like progress in the sensitivity of direct detection of halo dark-matter particles.

\section{Inner Space meet Outer Space}
The young field of particle physics and cosmology was nurtured at many places: numerous summer workshops at the Aspen Center for Physics, several longer programs at the Institute for Theoretical Physics (now KITP) at the University of California, Santa Barbara, and at summer and winter schools around the globe.  Here I focus on two important institutions that helped to spark the marriage and one important meeting that continues to this day.

\subsection{The mother church}
In the summer of 1981, my mentor David Schramm went for a hike in the Dolomites with Leon Lederman, Director of Fermilab.  They came back with the idea of an astrophysics group at Fermilab, and challenged Hans Mark, Deputy Administrator of NASA, to fund such a group at a particle physics lab.  He did.  Lederman recruited Rocky Kolb and me to lead the group, and it came into existence in the fall of 1983.  
Fermilab became the ``mother church'' for the coming together of inner space and outer space.  It served as the hub for this exciting, new activity:  graduate students and postdocs were trained, topical workshops were held, and the two cultures were mixed. 

Fermilab hosted the first major conference on particle physics and cosmology in May 1984 -- {\it Inner Space/Outer Space:  The Interface Between Cosmology and Particle Physics}.  There were more than 220 participants, drawn from astrophysics, cosmology, and particle physics.  Kolb and I convinced the scholarly University of Chicago Press, which didn't ordinarily publish conference proceedings, to publish the proceedings \cite{ISOS1984} by telling them it would mark the birth of a new field.  

Looking at the list of participants and the table of contents, {\it Inner Space/Outer Space} lived up to its hype.  Inflation, cold dark matter, magnetic monopoles, supersymmetry, the CMB and large-scale structure were all covered.  A paper by James Hartle was a prelude to the Hartle-Hawking wavefunction of the Universe.  The only thing missing was superstring theory -- the first string revolution was to come months later in Aspen, CO.  The proceedings also featured cartoons and other illustrations (see pages xi, 172, 286, 444, 446, and 494) and the conference had a T-shirt (shown on page 2), possibly a first for a science meeting.

The final speaker of the meeting was Steven Weinberg.\footnote{In December 1983 he organized the first Jerusalem Winter School and chose the topic, {\it The Intersection between elementary particle physics and cosmology}, and I was one of the lecturers.}  His 
lecture was on physics in higher dimensions, and it focused on Kaluza-Klein theories.  In the Q\&A, Weinberg was asked how Kaluza learned how to swim.  His answer:  ``... in the fashion of a true theorist:  he read a book on how to swim, jumped in the water, and swam.''

Among the other legacies of the NASA/Fermilab Astrophysics Center was the Sloan Digital Sky Survey (SDSS).  It transformed astronomy by pioneering survey science and was the forerunner for the Dark Energy Survey, the Dark Energy Spectroscopic Instrument (DESI), and most recently Rubin/LSST.  Fermilab played a central role in the SDSS, and its former Director John Peoples served as the SDSS Director.  The primary science driver for the SDSS was to map large-scale structure with a million-galaxy redshift survey, cf., 2400 galaxies for CfA$_1$, to test inflation and cold dark matter.  The five-year survey began in 1998, and follow on's to SDSS continue to this day.

\begin{figure}[tbp]
\center\includegraphics[width = 0.95\textwidth]{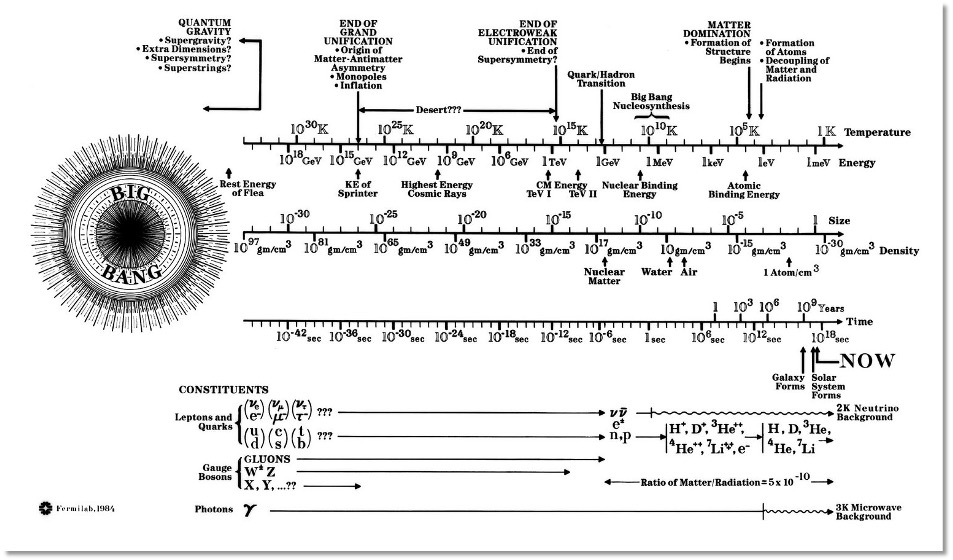}
\caption{The iconic Fermilab poster, the complete history of the Universe.  It was a collaboration between Angela Gonzales, the famous Fermilab artist/illustrator, and me.}
\label{Fermilab_Universe}     
\end{figure}

\begin{figure}
\centering
\begin{subfigure}{0.4\textwidth}
  \centering
  \includegraphics[width=0.9\linewidth]{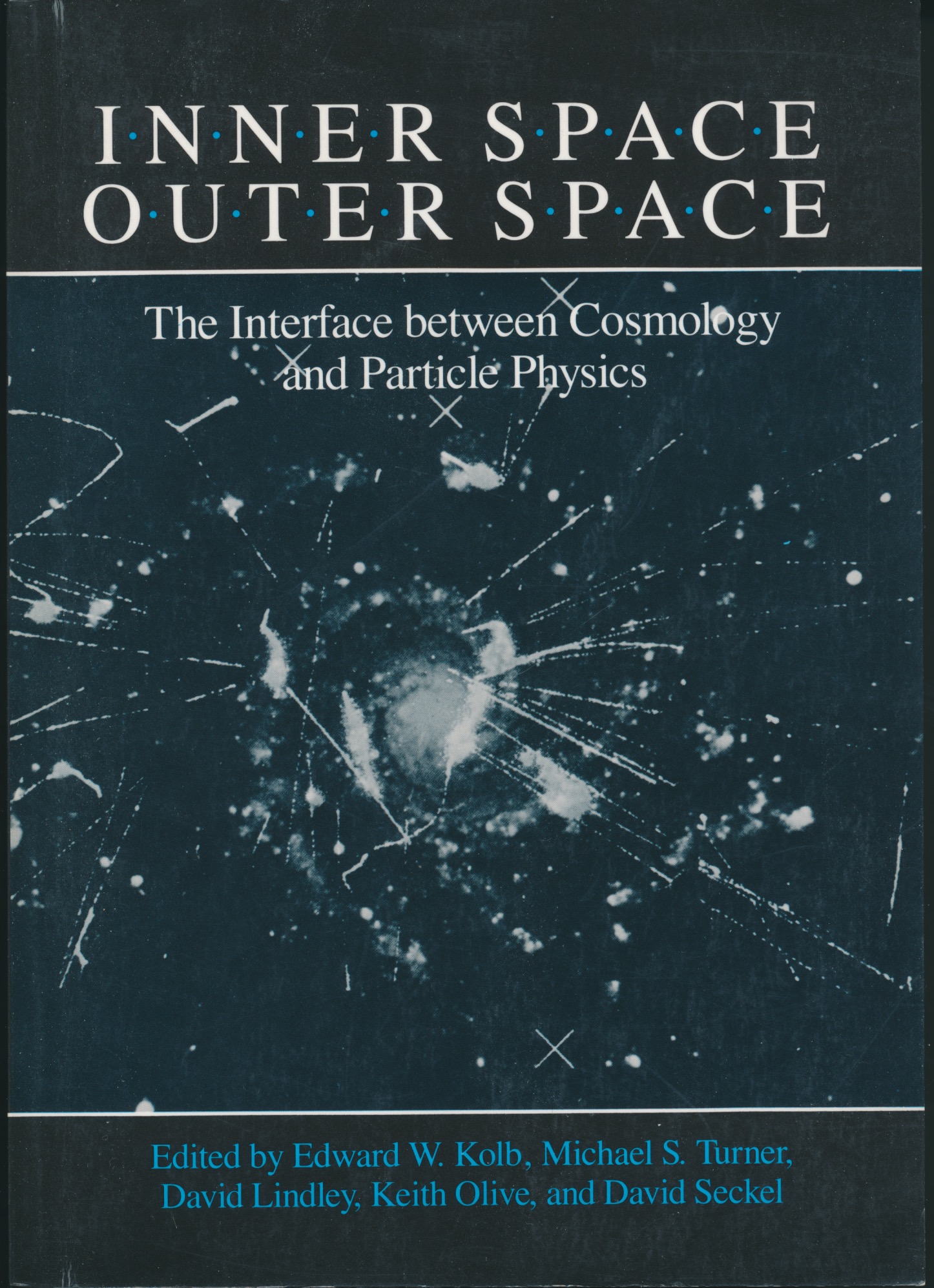}
  \caption{}
\end{subfigure}%
\begin{subfigure}{0.4\textwidth}
  \centering
  \includegraphics[width=0.9\linewidth]{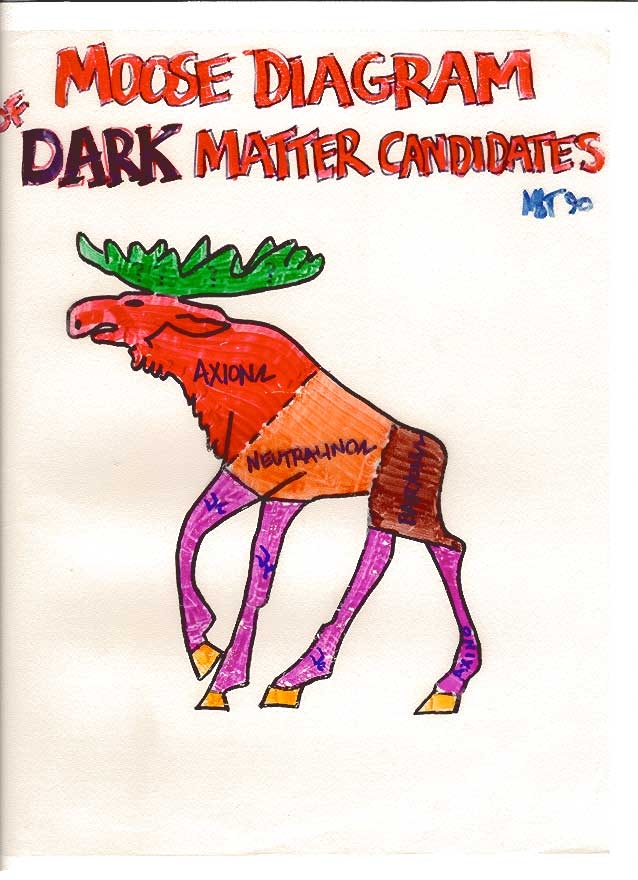}
 \caption{}
\end{subfigure}
\caption{(a):  Proceedings of the 1984 Inner Space/Outer Space meeting at Fermilab.  (b):  Dark matter candidates circa 1990 (my well-used transparency).}
\label{ISOS_Q2C}
\end{figure}

\subsection{Center for Particle Astrophysics}
In 1989 NSF Director Erich Bloch initiated a new centers program, the NSF Science and Technology Centers.  While it was not met with enthusiasm from many scientists, one of the centers in the first cohort of eleven helped to further nurture and cement the connections between particle physics and cosmology.  

The Center for Particle Astrophysics (CfPA) at the University of California, Berkeley was led by Bernard Sadoulet, a particle physicist turned cosmologist, and Marc Davis, an astronomer.  The CfPA existed from 1989 to 2001 and funded projects at the intersection of particle physics and cosmology as well as stimulating interactions between cosmologists and particle physicists.  

Its dark-matter portfolio included the design of and R\&D for low-temperature, low-background dark-matter detectors (leading to the CDMS experiment) and the MACHO\footnote{MACHO was an acronym created by Kim Griest in response to my term WIMP.} project, a search for Massive Astrophysical Compact Halo Objects (MACHO) through microlensing.  It also supported a host of CMB projects, the most notable was BOOMERanG, which in 2000 provided the first evidence that the Universe is flat \cite{BOOMERanG2000}, testing a key prediction inflation.

\subsection{Februaries in Los Angeles}
In 1994 UCLA began hosting a biennial meeting on dark matter. The driving force behind the meeting was particle experimentalist turned dark matter physicist David Cline. The first  meeting took place in Santa Monica in February 1994,\footnote{Though the proceedings listed the location as Bel Air, the tony LA neighborhood north of UCLA.} and the topics included dark-matter candidates, searches and constraints.  

The announcement of the accelerating Universe was made at the 1998 meeting, and thereafter the meeting was called, {\it International Symposium on Sources and Detection of Dark Matter and Dark Energy in the Universe}.  This influential gathering, strategically located in a warm location during the winter, has continued to attract both particle physicists and astronomers, and all those in between.

\section{Nonbaryonic dark matter}\label{baryons}
Establishing that dark matter is the dominant form of matter in the Universe involved showing that the gravity of stars is insufficient to hold galaxies and clusters of galaxies together.  And because stars contribute to little to the cosmic mass/energy budget, about 0.5\% of the critical density, to elevate that argument to the cosmic scale one need only show that the total mass density exceeds 0.5\% of critical.  By the 1980s there was already plenty of evidence that the total mass density was 10\% of critical density or greater.\footnote{Here and throughout, $\Omega_0$ is the fraction of critical density contributed by all forms of mass and energy today; for a flat (spatially uncurved) Universe, as favored by inflation, $\Omega_0=1$.  $\Omega_i$ is the fraction contribution by photons ($\sim 0.01\%$), stars ($\sim 0.5\%$), baryons, matter, dark energy and so on.  The critical density $\rho_{crit} = 3H_0^2/8\pi G = 1.88h^2 10^{-29}\,$g/cm$^3$, where $H_0 =100h\,$km/sec/Mpc.}

To show that dark matter is not made of baryons, one must establish that $\Omega_M > \Omega_B$. As the evidence for dark matter emerged in the 1980s, it was realized that determinations of the total matter density were all lower limits, because the full extent of galaxy halos had not been determined.  The best one could say was $\Omega_M > 0.1$ or $0.2$.\footnote{Determining the mean matter density is no ``mean'' feat:  In principle one needs to measure the mass of a well-defined and large enough region of space to be a fair sample and then divide by its volume.  Instead, estimates of the mass-to-light ratios of galaxies -- all lower limits -- were multiplied by the measured luminosity density \cite{FG1979}.}  As larger and larger volumes were probed, $\Omega_M$ seem to rise, but the evidence for a continued rise to $\Omega_0 = 1$ was weak at best.

 Big Bang Nucleosynthesis (BBN) provides a straightforward means of determining the baryon density \cite{Yangetal1984,Reeves1973}:  the production of deuterium in the big bang depends strongly upon the baryon density, D/H\,$\propto \Omega_B^{-1.7}$ and stellar processes since BBN only destroy deuterium.  For years, the measurement of D/H in the local interstellar medium, $D/H \simeq 2 \times 10^{-5}$, provided the best lower limit to the BBN production of deuterium, leading to the upper limit:  $\Omega_B < 0.04 h^{-2} < 0.14$ for $h>0.5$.  

Circa 1980, the data could be read two ways:  With $h\simeq 0.5$, $\Omega_0 \sim \Omega_M \sim \Omega_B \sim 0.1$ or, there is no need for nonbaryonic dark matter, or $\Omega_M\sim 0.2$ or greater and $\Omega_B \sim 0.1$ and there is a need for nonbaryonic dark matter.

While inflation provided strong motivation for $\Omega_0 =1$ -- at least for theorists -- the evidence was lacking until 2000.
However, in 1997 the primordial abundance of deuterium was measured by Burles and Tytler \cite{BT1997}, which pinned down $\Omega_Bh^2 = 0.0193\pm 0.0014$; in 1998 the acceleration of the expansion of the Universe and dark energy were discovered \cite{SN_teams}; and in 2000 the balloon-borne BOOMERanG experiment determined key cosmological parameters \cite{BOOMERanG2000,Bond2000} including $\Omega_0 = 1 \pm 0.06$, $\Omega_Mh^2 = 0.20 \pm 0.02$ and $\Omega_Bh^2 = 0.03\pm 0.005$.  In turn, this showed there was a statistically significant gap between the total matter density and that contributed by baryons.

Today, the evidence for nonbaryonic dark matter is greater than $50\sigma$:  
$$\Omega_Bh^2 = 0.0222\pm 0.0002 < \Omega_Mh^2 = 0.143 \pm 0.001,$$
where the baryon density is derived from two consistent determinations of equal precision, the CMB and BBN.  The matter density is derived from large-scale structure measurements and the CMB. 

The agreement between the two measurements of the baryon density based upon very different physics and at very different epochs is a poster child for precision cosmology.  And further, the discrepancy between the baryonic contribution and the total matter density makes nonbaryonic dark matter a powerful falsifiable prediction of the current cosmological paradigm.

\section{The rise of $\Lambda$CDM}
\subsection{COBE!}
Beginning in the mid 1980s inflation + CDM became the new paradigm for structure formation.  Data slowly accumulated to support it.  The detection of CMB anisotropy on angular scales around 10 degrees by the COBE satellite in 1992 \cite{COBE} was a major milestone and by my reckoning the birth of precision cosmology.\footnote{Equally impressive was the measurement of the blackbody spectrum (no deviations from a perfect blackbody larger than 0.03\%) and the temperature to 4 significant figures.}

Not only was the COBE detection the first evidence of CMB anisotropy beyond the kinematic dipole, but it was also consistent with the inflationary prediction and allowed the overall amplitude of the spectrum of density perturbations to be normalized.  

The COBE result led to a race to measure the anisotropy on even smaller angular scales (and higher multipole number), to test inflation + CDM and to precisely determine cosmological parameters, including $\Omega_M$ and $\Omega_B$.  A series of ground-based and balloon-borne experiments were undertaken and there were new results reported every few months.  Two satellite experiments were planned for launch after 2000:  the NASA's WMAP and ESA's Planck.

The predicted CMB multipole power spectrum is by now familiar:  a plateau at $\ell < 30$, followed by a series of acoustic peaks, with the position of the first peak being determined by $\Omega_0$ \cite{BE1987}.  For a flat Universe, the first peak occurs at $\ell \simeq 200$.  From 1992 to 2000, results for larger and large $\ell$s came in from different experiments and a peak at $\ell \sim 200$ ($\theta \sim 1^\circ$) began to appear.  As mentioned earlier, in 2000 BOOMERanG balloon borne instrument resolved the first peak by measuring the power spectrum out to $\ell \simeq 600$ \cite{BOOMERanG2000}.

\subsection{The $\Omega$ problem}
Even as the inflation + CDM paradigm gained momentum, there was one very nagging issue:  $\Omega_M$ was still no where near 1.  On the other hand, measurements of the mean matter density still did not extend to large enough scales to get a fair sample of the Universe. Optimist that I am, I was worried enough that I wrote a paper about the $\Omega$-problem, explaining how to have a flat Universe and $\Omega_M \sim 0.1 - 0.3$ \cite{TSK1984}, with either vacuum energy ($\Lambda$) or decaying dark matter.

There were other problems as well:  The COBE normalization of the spectrum of perturbations and large-scale structure measurements preferred $\Omega_Mh \sim 0.3$, which if $\Omega_M = 1$, implied a very low Hubble constant.  And then there was the Hubble constant-age tension:  for $\Omega_M =1$, the age of the Universe $t_0 = 2H_0^{-1}/3 \simeq 6.5h^{-1}\,$Gyr.  While the Hubble constant was still poorly determined -- between 50\,km/s/Mpc and 100\,km/s/Mpc -- unless it was very close to 50\,km/s/Mpc, the Universe was in danger of being younger than its oldest stars, argued to be 10 to 14 Gyr old.

For me, it all came to a head in 1993, with a powerful argument that used the baryon density as determined by BBN and the ratio of the total mass of a cluster (determined by the virial theorem) to its baryonic mass (as determined by the x-ray emission from hot cluster gas where most of the baryons reside) to infer $\Omega_M \simeq 0.3$ \cite{WNEF1993}.  Though clusters are rare objects, they are large enough to provide a fair sample of the Universe.  Any wiggle room for $\Omega_M = 1$ was rapidly disappearing.

\subsection{Cosmic Acceleration:  predicted and detected}
In 1995, Krauss and I wrote a paper entitled \cite{KT1995}, {\it The cosmological constant is back}, showing that all the problems of the inflation + CDM paradigm -- age/Hubble, cluster baryon fraction, large-scale structure -- could be solved by adding a cosmological constant ($\Omega_M \sim 0.3, \Omega_\Lambda \sim 0.7, h \sim 0.7$).\footnote{A few months later, a similar proposal was made by Ostriker and Steinhardt \cite{OS1995}.}  

$\Lambda$ had a very checkered history in cosmology, having been trotted out whenever cosmology had a problem going back to Einstein \cite{FTH2008}, and an even worse reputation in particle physics because of the vacuum energy problem \cite{WeinbergRMP}.  I remember being surprised that our proposal received a mildly positive reception, though it only received the third prize in the 1995 Gravity Research Foundation's essay competition.

Other solutions were suggested as well, e.g., open inflation, a flat Universe with $H_0 = 30\,$km/s/Mpc, and hot + cold dark matter.  For all its warts, $\Lambda$CDM worked the best \cite{DGT1996,Dialogues}, and made a striking prediction:  the Universe should be speeding up, not slowing down.

It all came together very quickly.  In 1998, the two teams using type Ia supernovae to determine how fast the expansion of the Universe was slowing down discovered that it was actually speeding up \cite{SN_teams}; in 2000, BOOMERanG \cite{BOOMERanG2000} showed the Universe was flat consistent with $\Lambda$ and that $\Omega_M$ was significantly greater than $\Omega_B$; and the HST Key Project announced its results, $H_0 = 71 \pm 6\,$km/s/Mpc \cite{H0_Key_Project}.  $\Lambda$CDM was established with its mysterious dark energy and nonbaryonic dark matter:  $\Omega_0 = 1.08 \pm 0.06 $, $\Omega_\Lambda = 0.66 \pm 0.06$, $\Omega_M = 0.33 \pm 0.06$ and $\Omega_B = 0.05 \pm 0.01$ \cite{Bond2000}.

\section{Coda}

\subsection{Rich theoretical speculation}
The period of 1980 to 2000 was characterized by much theoretical speculation and exploration of the deep connections between inner space and outer space. 
Particle dark matter was a central bridge between the two disciplines, and is even more so today.  I  briefly discussed inflation (covered in more depth in the contribution of Guth) and the rise of $\Lambda$CDM because of their deep connections with particle dark matter.  

There are many dark matter candidates which I haven't discussed, e.g., sterile neutrinos, a warm dark matter candidate \cite{DodelsonWidrow}, the gravitational production of particles during inflation, both light \cite{TurnerWidrow} and heavy \cite{KolbChungRiotto}, the briefly-lived 17 keV neutrino, and superheavy magnetic monopoles, a whole subject in and of itself. 

And there were other intriguing ideas about dark matter and related topics from this period that haven't panned out, at least not yet.  They include cosmic strings, MOdified Newtonian Dynamics (MOND), MACHOs, and the DAMA/Libra annual modulation.  Each had its own part in the particle dark matter story, for which there is not enough space here to properly describe.

\subsection{Triumphs}
\begin{figure}
\centering
\begin{subfigure}{0.27\textwidth}
  \centering
  \includegraphics[width=0.90\linewidth]{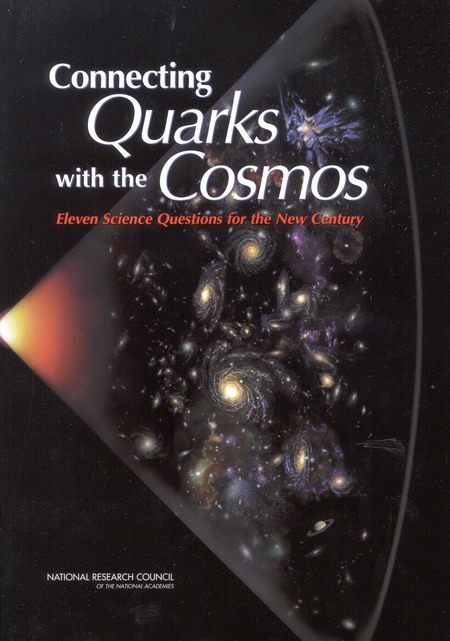}
  \caption{}
\end{subfigure}%
\begin{subfigure}{0.73\textwidth}
  \centering
  \includegraphics[width=0.95\linewidth]{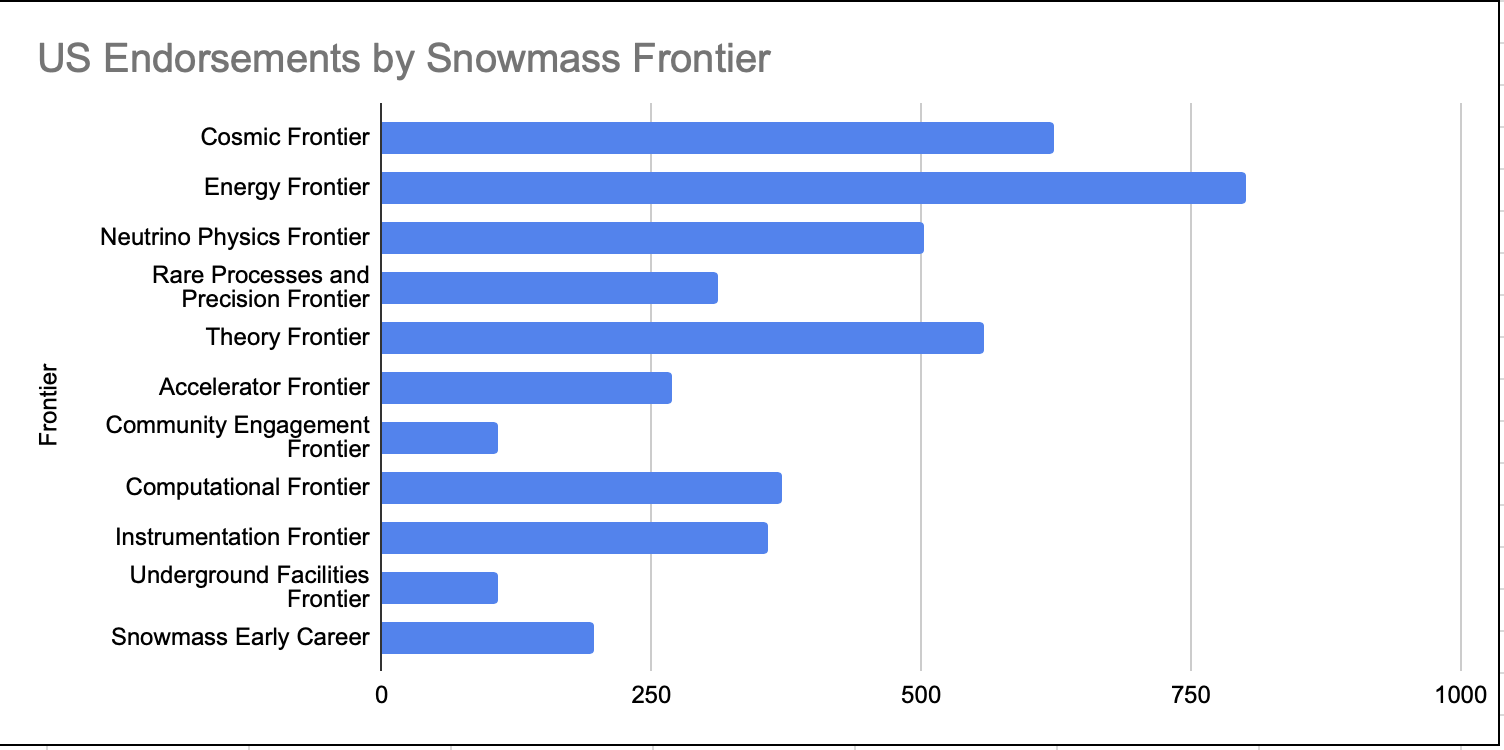}
 \caption{}
\end{subfigure}
\caption{(a):  2000 NAS study, Connecting Quarks with the Cosmos.  (b):  Endorsements of the 2023 P5 report by particle physics frontier (Snowmass 2023).}
\label{Q2C}
\end{figure}

The period 1980 to 1990 transformed cosmology.  Entering the period, the Universe we understood was made of stars and studied by astronomers.  By the end of the period, stars were a minor component in a Universe dominated by dark matter and dark energy, with a beginning whose story involved fundamental physics and whose events shaped all that we see today.
In 1980, particle physics and cosmology were two separate fields, and by 2000 they were deeply connected through their agendas and their participants; see for example, Fig.~\ref{Q2C}.  

This period also saw the beginning of precision cosmology, in a field where the errors used to be in the exponents.  Here are some of the numbers that today precisely characterize the Universe \cite{Planck2018}: $T_0 = 2.7255 \pm 0.0006\,$K, 
$\Omega_0 = 1.001 \pm 0.002$, $\Omega_\Lambda = 0.69 \pm 0.006$, $\Omega_M = 0.31 \pm 0.006$, $\Omega_B = 0.049 \pm 0.001$, $H_0 = 70 \pm 3\,$km/s/Mpc,\footnote{The Hubble constant is once again front and center, which my $70 \pm 3$ is hiding.  There is significant tension between the direct determinations of $H_0$ and the indirect determinations involving the CMB \cite{HubbleTension}.  Whether or not this leads to new insights reminds to be seen.} and $t_0 = 13.79 \pm 0.02\,$Gyr.

The advances in cosmology parallel those in particle physics during this time, shifted by 20 years or so.  In particle physics, there was the rise and establishment of the Standard Model to high precision; in cosmology, it was $\Lambda$CDM and precision cosmology.  The 21st century as been devoted to looking for the cracks in the Standard Model and $\Lambda$CDM which will lead to the next paradigm and an even deeper understanding of the Universe and the basic rules that govern matter, energy, space and time.

\subsection{Challenges}

Both standard models are great achievements, but neither cosmology nor particle physics are solved.  Instead, they are left with bigger and more profound questions to answer, and big ideas and powerful tools to do so.  That said, the path forward in both fields has been slower than desired.  But isn't that always the case?

Here are a few of the big questions that connect the two fields,
\begin{itemize}
    \item What is the dark matter particle(s) and where does it fit into to the grander theory of particles and forces?
    \item What is the nature of dark energy and why is the energy of the vacuum so small?
    \item Why is the ratio of particle dark matter to baryonic matter around 5 and not much larger or smaller \cite{TurnerCarr1986}?  How are the origins of baryonic matter and particle dark matter related?
    \item What is the origin of space, time and the Universe?
\end{itemize}

One might be tempted to call the period of 1980 to 2000 a Golden Age; however, according to Greek mythology a Golden Age is followed by periods of decline.  So I won't refer to it as a Golden Age, because I believe the best is still yet to come for particle physics and cosmology.

\subsection{Endnote}
Everyone sees history from their own personal perspective -- the Rashomon effect -- and I am no different.  For this reason I have included my worldline during the period discussed as an appendix to provide some context for where I saw things from.  My talk at the Symposium and this write up benefited from previous work, e.g., the review I wrote on the path to precision cosmology and $\Lambda$CDM \cite{Turner2022}, as well as two other meetings I participated in.  The first was the 2018 conference on the Physics of Galaxy Scaling Relations and the Nature of Dark Matter, held at Queen’s University in 2018, whose panel discussion is available \cite{panel}, and the second is the Vera Rubin Symposium at Georgetown University in June 2019 \cite{Georgetown}.  Finally, I mention another paper that addresses the history of particle dark matter by Bertone and Hooper \cite{BertoneHooper}.  Their review has the advantage of the authors not being participants in the activities of 1980 to 2000.

\vskip 0.2in
\noindent{I thank the organizers and participants of the 4th International Symposium on the History of Particle Physics held at CERN, 10 - 13 November 2025.  The meeting was both lively and enjoyable.}

\section*{Appendix:  My worldline 1980 -- 2000}
\begin{itemize}
    \item 1971 -- 1973:  Particle physics graduate student at SLAC supervised by Fred Gilman (supported by an NSF Fellowship)
    \item 1973 -- 1975:  Lab animal caretaker at Stanford Hospital and self-employed auto mechanic (on leave from Stanford)
    \item 1975 -- 1978:  Graduate student at Stanford supervised by Robert Wagoner (1978 PhD thesis on gravitational waves)
    \item 1978 -- 1980:  Enrico Fermi Fellow UChicago mentored by David Schramm (BBN, baryogenesis and early Universe cosmology)
    \item 1980:  Appointed Assistant Professor UChicago (Department of Astronomy \& Astrophysics and the Enrico Fermi Institute)
    \item 1980 -- 1981:  On leave from Chicago as one of the first postdoctoral fellows at ITP UCSB (early Universe cosmology program)
    \item 1982:  Nuffield Workshop at Cambridge University (Bardeen, Steinhardt and Turner paper on the origin of density perturbations)
    \item 1983 -- 1997:  Co-founder and co-head (with Edward Kolb) of the NASA/Fermilab Astrophysics Center
    \item 1983:  Associate Professor (with tenure) of Physics and of Astronomy \& Astrophysics UChicago (50\% Fermilab, 50\% UChicago)
    \item 1984:  Inner Space/Outer Space Conference at Fermilab (coming out party for particle cosmology)
    \item 1985:  IAU Symposium 117:  Dark Matter in the Universe (Dark Matter Candidates)
    \item 1988:  Co-organizer of the ITP UCSB Microphysical Cosmology Program 
    \item 1988:  CERN/ESO Symposium on Astrophysics, Cosmology and Fundamental Physics in Bologna (Dark Matter Candidates)
    \item 1990 Publication of {\it The Early Universe} (Kolb and Turner)
    \item 1990, 1998, 2003:  Nobel Symposia 79, 109, and 127 (Particle Physics and Cosmology)
    \item 1991 First visit to CERN (Windows on the Axion)
    \item 1994:  Sabbatical at Center for Particle Astrophysics UCBerkeley
    \item 1996:  Critical Dialogues in Cosmology at Princeton ($\Lambda$CDM wins the CDM competition)
    \item 1997:  Election to the NAS, Chair of Astronomy \& Astrophysics UChicago (1997 to 2003), and death of mentor David N. Schramm
    \item 2000 -- 2002:  Chair, Physics of the Universe Committee ({\it Connecting Quarks with the Cosmos} report)
    
\end{itemize}



\end{document}